\documentclass[a4paper]{JAC2003}
\usepackage{graphicx}
\usepackage{booktabs}
\usepackage{varwidth}
\usepackage{xcolor}
\begin{document}
\title{SIMULATION OF BEAM--BEAM INDUCED EMITTANCE GROWTH IN THE HL-LHC WITH CRAB CAVITIES\thanks{This work was partially supported by the US LHC Accelerator Research Program and the National Energy Research Scientific Computing Center of the US Department of Energy under contract No. DE-AC02-05CH11231}}

\author{S. Paret, J. Qiang, LBNL, Berkeley, CA, USA}

\maketitle

\begin{abstract}
The emittance growth in the HL-LHC due to beam--beam effects is examined by virtue of strong--strong computer simulations. A model of the transverse damper and the noise level have been tuned to simulate the emittance growth in the present LHC. Simulations with projected HL-LHC beam parameters and crab cavities are discussed. It is shown that with the nominal working point, the large beam--beam tune shift moves the beam into a resonance that causes substantial emittance growth. Increasing the working point slightly is demonstrated to be very beneficial.
\end{abstract}

\section{Introduction}
The force between two colliding beams applies a coherent kick to the colliding bunches if they have a finite offset.
At the same time, due to its non-linear nature, the beam--beam force damps coherent transverse motion at the cost of the emittance.
Noise on the transverse bunch positions at Interaction Points (IPs) can therefore lead to emittance growth.
This effect is more pronounced for higher beam intensity and therefore of particular interest for the
High-Luminosity Large Hadron Collider (HL-LHC).

In addition to the extreme beam parameters, a new feature of the HL-LHC may impact the emittance.
The HL-LHC layout is based on Crab Cavities (CCs) to compensate the geometric luminosity loss due to large crossing angles.
Large crossing angles are required to mitigate long-range beam--beam effects.
Noise in the phase of the CCs' field imparts a transverse offset on to the colliding bunches.
Hence noise in the CCs may accelerate the emittance growth~\cite{ohm01}.
A prediction of the emittance growth, depending on the noise, is of considerable interest for HL-LHC planning.

Simulations to predict the impact of CC noise on the emittance have been carried out for years (see, e.\,g., Refs.~\cite{ohm01,cal01,qia01}).
But over time the anticipated beam parameters in the future LHC have changed, and recently a detailed model of the LHC's transverse damper has been implemented in the code BeamBeam3D~\cite{par01}.
The damper has a significant impact on the evolution of the emittance, because it suppresses coherent dipolar motion -- ideally without contributing to the emittance growth.
However, in reality, noise in the damper -- in particular, due to the uncertainty of the beam position measurements -- has a detrimental effect on the emittance.
The measurement uncertainty is included in the damper model.

Since the noise properties of the future CCs and their control system are not known yet, reasonable models and parameter ranges have to be explored.
Common noise models are white noise, coloured noise with a specific correlation time, and a mono-frequent perturbation~\cite{ohm02}.
White noise is the easiest to model and has only one free parameter, the r.m.s.\@ amplitude, but tends to be too pessimistic.
Correlated noise features an additional parameter, the correlation time.
A sinusoidal perturbation is described by an amplitude, frequency, and phase.
In addition, the phase relation between the perturbation on a CC before and after an IP (in a local crabbing scheme) is assumed to play a role.
In this paper, only white CC phase noise is considered.
The examination of the other noises is work in progress.

The set-up of the simulations is described in Section~1.
The first case that we consider here is a LHC run from last year.
The set of measured beam parameters was used in simulations to reproduce the measured emittance growth.
The purpose of this study was to validate the code and to determine the noise level in the damper, which is not precisely known.
Assuming that the damper hardware does not change, a similar noise level is expected to be present in the HL-LHC.
Section~2 describes the simulation of the recent LHC performance.
In the following section, we examine the impact of CCs on last year's beam.

The focus then turns towards the HL-LHC.
As yet, a definite plan for the HL-LHC set-up does not exist.
Two HL operational scenarios based on different bunch spacing have been specified~\cite{bru01},
and although luminosity levelling is a key element of both HL-LHC scenarios, the means to achieve it have not yet been defined.
As a consequence, only studies for possible HL-LHC conditions can be run at this point.
Section 4 is dedicated to various case studies with HL beam parameters, mainly referring to the 50\,ns bunch spacing scenario.
The paper closes with a conclusion and an outlook.

\section{Computational set-up}
All of the simulations presented in this paper were done using a strong--strong collision model implemented in the code BeamBeam3D~\cite{qia02}.
In order to avoid numerically induced emittance growth, and to gain computation speed, the fields were computed assuming a Gaussian particle distribution instead of a self-consistent approach~\cite{par01}.
This assumption is justified by the fact that the particle distribution is Gaussian initially and does not change significantly in a short period of time under stable conditions.

The main objective of the simulations was to quantify the beam--beam induced emittance growth.
As shown in the next section, the emittance in the LHC increased in the order of 10\,\%/h in operation in the year 2012, which results in very small changes in a simulation that is limited to a few seconds of real storage time.
In order to keep the residual noise level low, $8 \times 10^6$ macro-particles were used.
The longitudinal space was discretized into eight slices.

The nominal tunes, $Q_x=64.31$ and $Q_y=59.32$, were initially set for the simulation of the present LHC as well as the HL-LHC.
Linear transfer maps, calculated using the working point, were employed to transfer the beams between collisions.
Two collisions per turn, corresponding to IP1 and IP5 in the LHC, were simulated.
Following the original, the crossing plane was horizontal in one IP and vertical in the other IP.
BeamBeam3D was modified to allow for changing collision planes, with the CC kicks and CC noise being applied correspondingly.

\begin{figure}[t]
  \centering
   \includegraphics*[width=70mm]{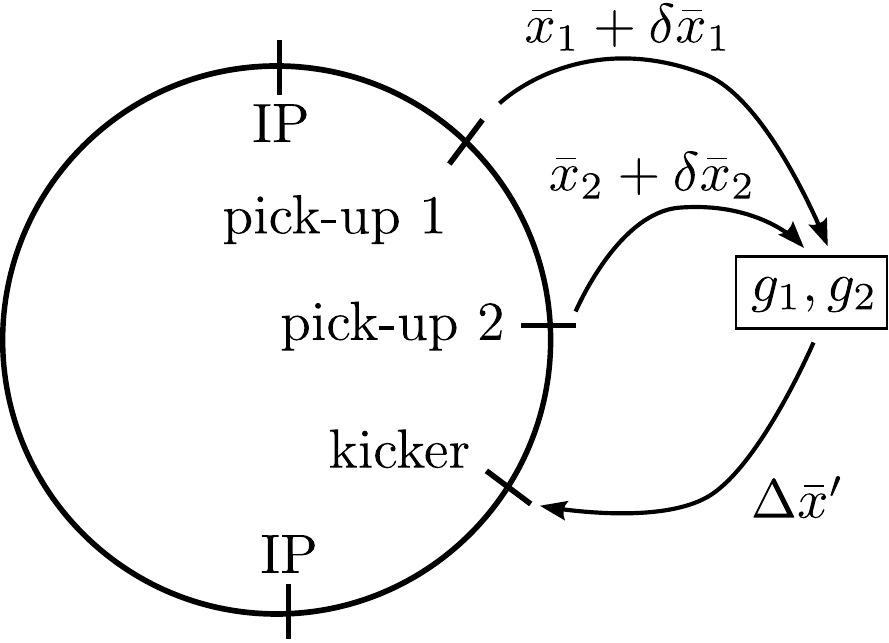}
  \caption{The scheme of the damper model.}
  \label{fig:scheme}
\end{figure}

The damper model uses a Hilbert-notch filter and two pick-ups per beam and plane, just as does the actual system in the LHC~\cite{hoe02,zha01}.
The correction kick at turn $n$ due to one pick-up is given by
\begin{equation}\label{eq:dx}
  \Delta \bar{x}_{n}' = \frac{a_0 g}{\sqrt{\beta_p \beta_k}} \sum^7_{m=1} H_m(\varphi_H) \times (\bar{x}_{n-d+1-m}-\bar{x}_{n-d-m}),
\end{equation}
where $H_m$ are the coefficients of the Hilbert filter, $\varphi$ is a phase that needs to be determined as a function of the tune and damper gain, and $d$ is the delay of the damper.
The actual kick is the superposition of two terms associated with different pick-ups.
Authentic values were used for the phase advance between the pick-ups and the kicker, and the phases in the Hilbert filter $\varphi_H$.
The gain of the damper was set to 0.1.
Noise is inserted by adding random numbers, $\delta \bar{x}_n$, to the measurement; that is, replacing $\bar{x}_n \rightarrow \bar{x}_n+\delta \bar{x}_n$ in Eq.~\ref{eq:dx}.
A scheme of the damping system is shown in Fig.~\ref{fig:scheme}.
In BeamBeam3D, the damper noise has a white spectrum and a Gaussian amplitude distribution. 

\section{The 2012 beam}
During regular operation, the luminous region in the detectors is measured, as well as the beam intensity.
The LHC operators provided a set of beam parameters from a `normal' long fill in June 2012 (fill 2710)~\cite{tra01}.
These parameters are only coarsely representative for the general LHC performance.
Notable variations of the emittance evolution and other parameters were observed in different fills.
However, detailed information about the overall performance is not yet available.

Assuming two equal bunches with Gaussian profiles colliding head-on, the beam width was calculated using
\begin{equation}
  \sigma_x = \sqrt{2} \cos \frac{\phi}{2} \sigma_{Lx},
\end{equation}
where $\phi$ is the crossing angle and $\sigma_{Lx}$ is the width of the luminous region.
The initial half cross-section of the beams found in that way is about 18\,$\mu m$ in the horizontal and vertical directions.
(The actual difference between the horizontal and vertical beam sizes was neglected.)
With the beta function at the IP, $\beta^* \!=\! 0.6$\,m, the emittance was deduced from the measured beam size.
Table~\ref{tab:params} lists the beam parameters used in the simulation of the LHC in 2012.
\begin{table}[t]
  \centering
  \caption{Beam parameters in the simulation of the LHC in 2012 and in the future.}
  \begin{tabular}{lcc}
        \hline\hline{}					&{\textbf{2012}}&{\textbf{HL}}\\
		\hline
    $N$  & $1.5\times10^{11}$ & $3.5\times10^{11}$ \\  
    $\epsilon_n$ [$\mu$m] & 2.3 & 3.0 \\  
    $\beta^*$ [m] & 0.6 & 1.02 \\  
    $Q_x$ & 64.31 & 64.31 \\  
    $Q_y$ & 59.32 & 59.32 \\ 
    $\theta$ [mrad] & 0.29 & 0.59 \\  
    $g_1+g_2$ & 0.1 & 0.1 \\  
    $f_{CC}$ [MHz] & - & 400.8 \\  
    Collisions per turn & 1 hor., 1 ver. & 1 hor., 1 ver. \\
       \hline\hline
  \end{tabular}
  \label{tab:params}
\end{table}

\begin{figure}[ht]
   \centering
   \includegraphics*[width=80mm]{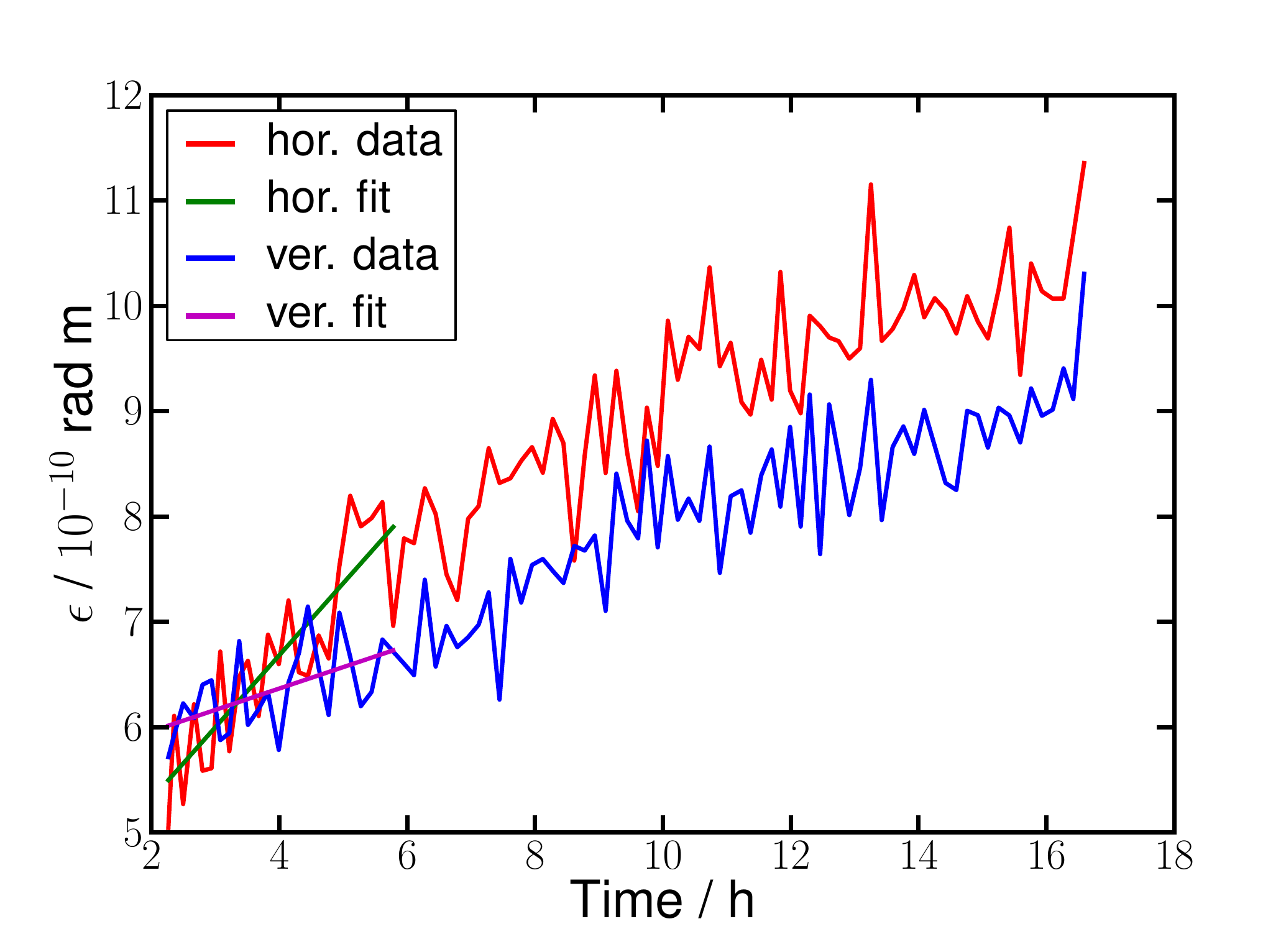}
   \caption{Transverse emittances in the LHC during a physics run in 2012.}
   \label{fig:emit_meas}
\end{figure}

Due to the computational cost, only a few tens of seconds of storage time can be simulated with more than a million macro-particles.
Hence only the initial emittance growth of the hour-long storage is of interest.
Due to the limited time resolution of the measurements, a linear fit to the emittance during the first 6 h after injection was performed to assess the initial growth.
Figure~\ref{fig:emit_meas} shows that the emittance growth in the horizontal and vertical plane is linear for several hours to a very good approximation.
The growth rates according to the fits are roughly 13\,\%/h in the horizontal plane and 4\,\%/h in the vertical plane.
Here and in what follows, we provide the average emittance of the two beams in either plane, because they are usually quite similar.

In the horizontal plane, intra-beam scattering contributes significantly to the emittance growth.
Simulations have been carried out to quantify its impact on the beams~\cite{sca01}.
For conditions similar to the ones considered here, an intra-beam scattering driven, horizontal emittance growth of about 5\,\%/h was found.
Therefore, we have assumed that the emittance growth due to the collisions amounts to about 8\,\%/h.

\begin{figure}[b]
   \centering
   \includegraphics*[width=80mm]{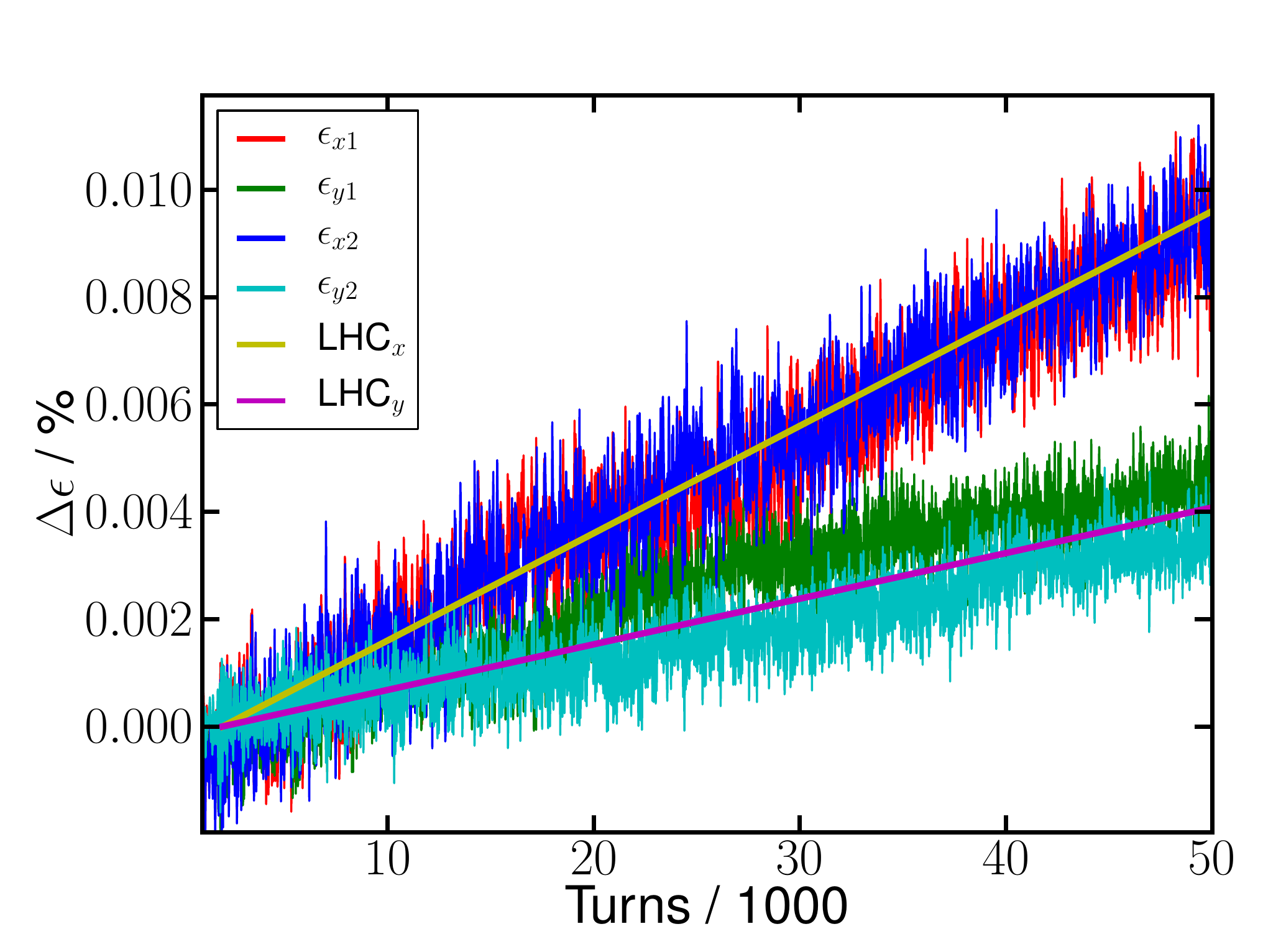}
   \caption{The emittance growth in a simulation of the LHC as operated in 2012. The straight solid lines visualize the fit to the measured data (Fig.~\ref{fig:emit_meas}) after correction for intra-beam scattering.}
   \label{fig:emit_ref}
\end{figure}

Simulations with different noise levels were run in order to reproduce the emittance growth attributed to the collisions with the actual beam and machine parameters.
The damper noise was adjusted to match the measured emittance growth.
Figure~\ref{fig:emit_ref} displays the simulated emittance that approximates the measured data.
The emittance of both beams is shown for both planes.
Two straight thick lines indicate the slope corresponding to 8\,\%/h and 5\,\%/h, respectively.
The r.m.s.\@ fluctuation of the beam centroid amounts 0.11\,$\mu m$ and 0.09\,$\mu m$, horizontally and vertically respectively, at an IP.
This fluctuation level is on the scale of observations in the LHC~\cite{hoe01}.

At this point, we have demonstrated that BeamBeam3D is able to reproduce the measured data of actual LHC beams.
Assuming that the damper noise does not depend on beam parameters, the r.m.s.\@ noise level fed into the damper in this simulation was used as a reference in other simulations.
The impact of crab cavities on the emittance of the LHC beams in 2012 is studied next.

\section{The 2012 beam with crab cavities}
Before switching to the HL-LHC parameters, crab cavities were virtually added to the 2012 LHC.
The set-up that has  been measured and simulated provides a good reference with which to compare the simulation with CCs.
For the HL parameters, it will take years before experimental data will be available, so the impact of CCs on HL beams can only be studied by comparing simulations with and without crab cavities.

In the first run, the beam and general machine parameters were kept as described in the previous section.
As the only change, CCs were added around both IPs, with to a phase advance of $\frac{\pi}{2}$ between the CC and the IP.
The CCs were assumed to be perfect; that is, the only noise source taken into account was the damper system.
The resulting emittance growth differs only weakly from the case without CCs.
Horizontally, we find 9\,\%/h and vertically only 2.4\,\%/h.

In order to get a first impression of the impact of CC noise on the emittance, a simulation without damper noise but with CC noise was done.
Since the noise in the future CCs is not known, the accelerating cavities in the LHC were taken for an estimation.
The power spectrum of the phase noise in these cavities has been measured and used to assess an approximate white noise level.
The white noise that contains the same power in all betatron sidebands as the actual spectrum corresponds to $2 \times 10^{-4}$ rad r.m.s.\@~\cite{bau01}.

\begin{figure}[b]
   \centering
   \includegraphics*[width=80mm]{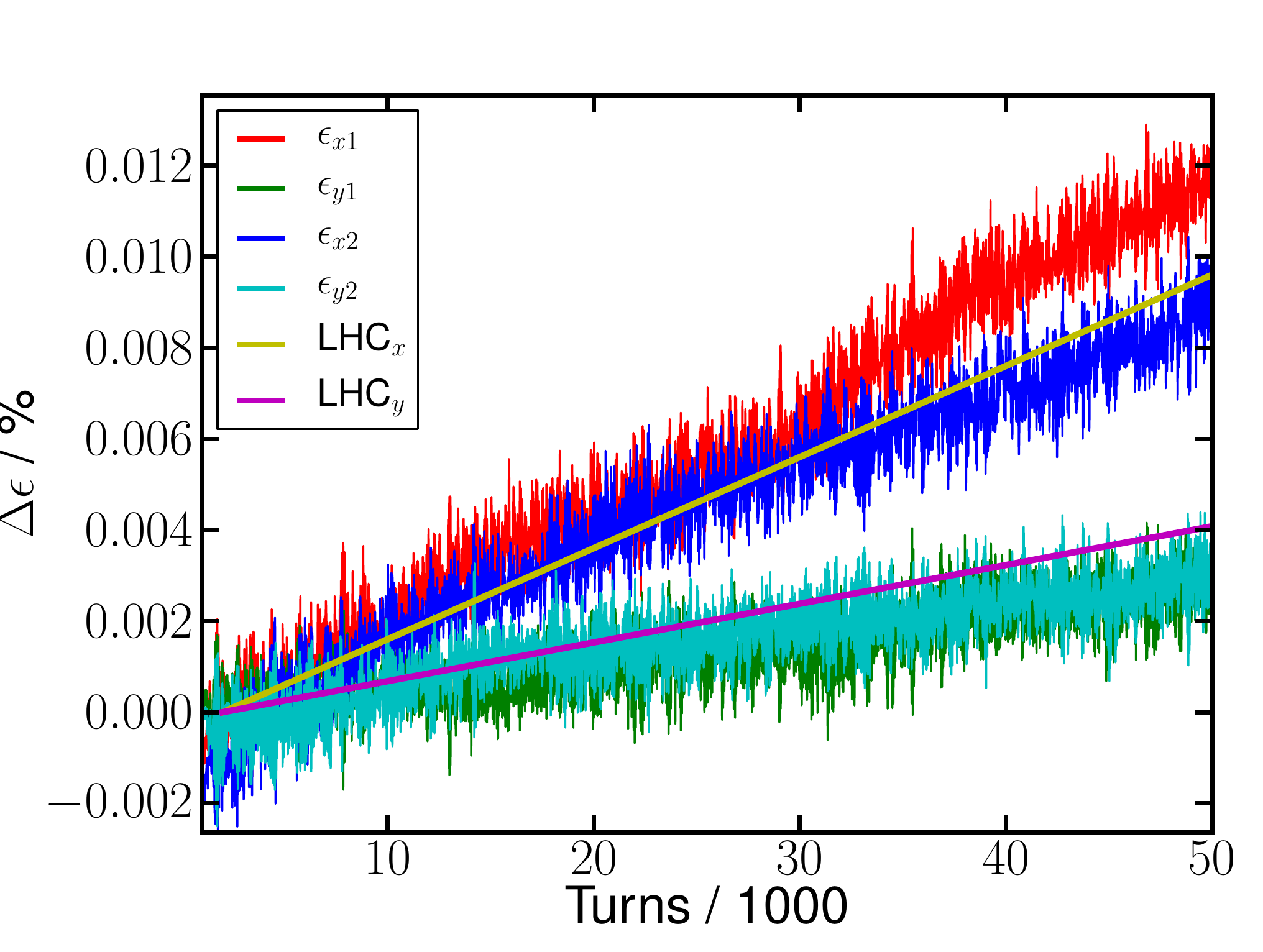}
   \caption{Emittances with the last 2012 beam parameters and CCs with damper noise only.}
   \label{fig:2012CCdamp}
\end{figure}

\begin{figure}[t]
   \centering
   \includegraphics*[width=80mm]{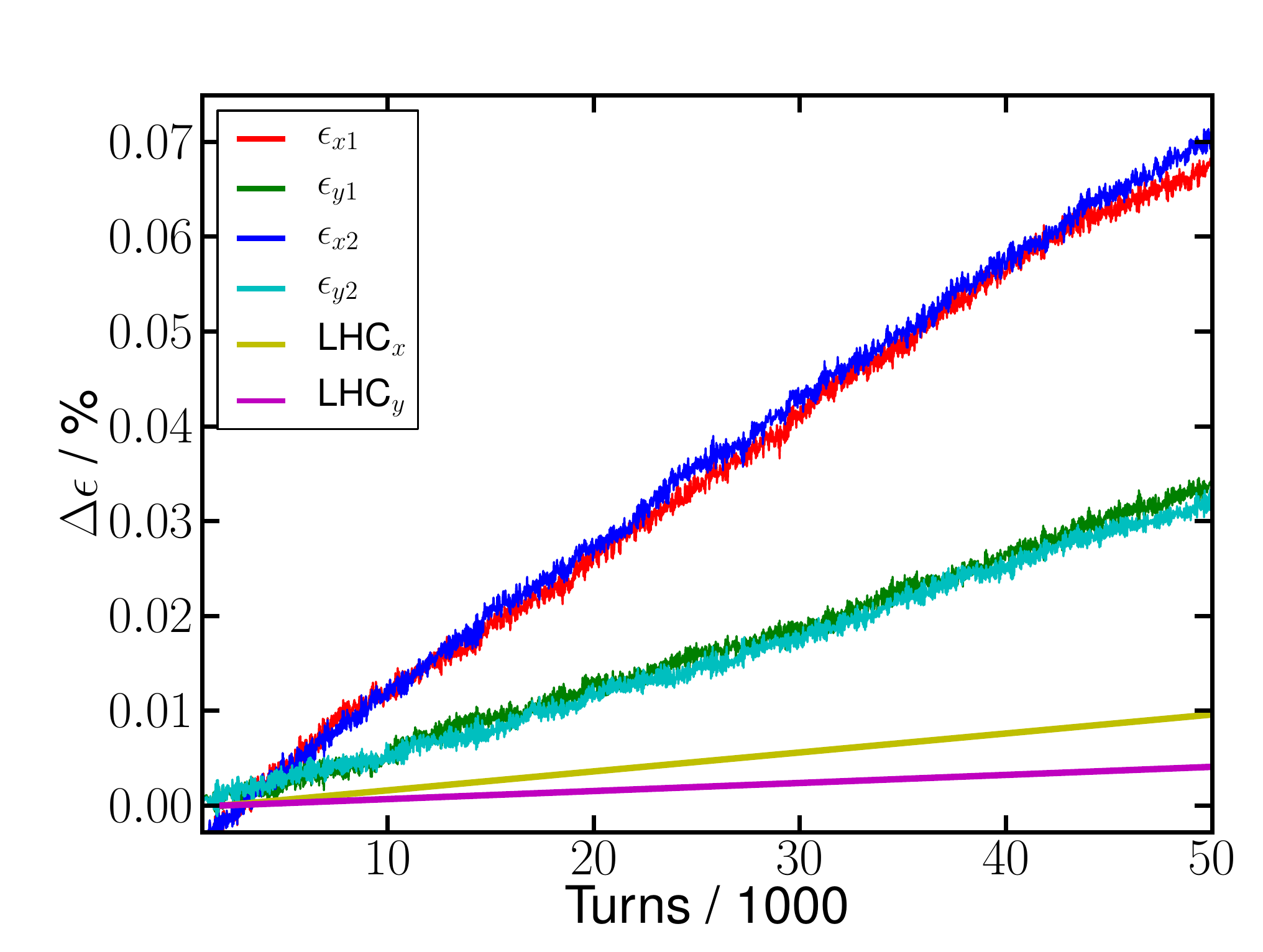}
   \caption{Emittances with the last 2012 beam parameters and CCs with CC noise only.}
   \label{fig:2012CCcrab}
\end{figure}

The evolution of the emittance with damper noise or CC noise is shown in Figs.~\ref{fig:2012CCdamp} and~\ref{fig:2012CCcrab}, respectively.
As far as damper noise is concerned, the CCs have little effect on the emittance.
Phase noise with the estimated level, on the other hand, has a severe impact on the emittance, increasing the growth to 60\,\%/h horizontally and 17\,\%/h vertically.

The emittance growth simulated with white noise is not an accurate prediction for the perturbation caused by CCs, however, for several reasons.
First, in the real system, no white noise is expected.
Simulations with more realistic noise spectra should give more accurate results.
A spectrum with lower noise at higher frequencies is expected to produce less emittance growth.
Second, the noise in the present accelerating cavities is only an upper boundary for the expected noise level.
Third, a filter to suppress the noise at the betatron sidebands is foreseen to further reduce the perturbation of the beam~\cite{hoe01}.

What we have learned so far is that even at a moderate beam intensity and crossing angle, the emittance is very sensitive to noise.
Noise in the damper plays only a marginal role.
In the next section, the same noise will be applied to HL beams.

\section{HL-LHC beams}
Now the HL beam parameters and noise are examined.
At present, two HL-LHC scenarios are considered~\cite{bru01}.
The primary difference is the number of bunches stored in the LHC.
The other beam parameters are adapted to match the luminosity goal and take other constraints, such as mitigation of long-range effects, into account.
Here, we focus on the 50\,ns bunch spacing option (with one exception).
The relevant beam parameters for our simulations are summarized in Table~\ref{tab:params}.

A key feature of the HL-LHC is luminosity levelling; that is, the maintenance of a certain luminosity, well below the achievable peak luminosity, for as long as possible~\cite{bru01}.
Different approaches to achieve levelling are under consideration, and they may affect the emittance in different ways~\cite{mur01}.
In this paper, we consider levelling by means of $\beta^*$, and therefore we have increased it from 15\,cm (the reference value in~\cite{bru01}) to $1.02$\,m.
Due to the small absolute emittance growth within the short time span covered by our simulations, an adjustment of $\beta^*$ to compensate the luminosity loss due to the emittance growth is not necessary.

\begin{figure}[t]
  \centering
  \includegraphics*[width=80mm]{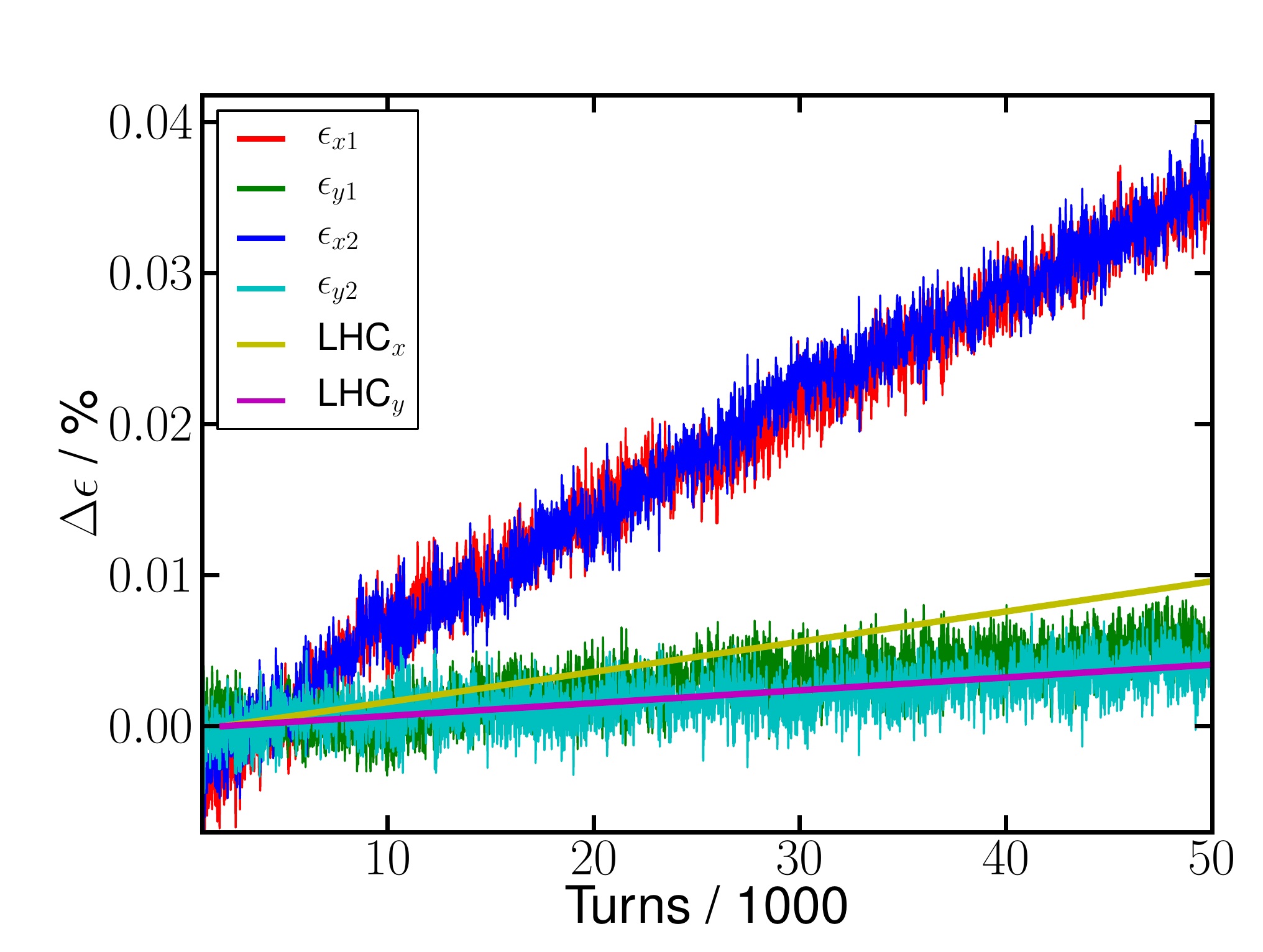}
  \includegraphics*[width=80mm]{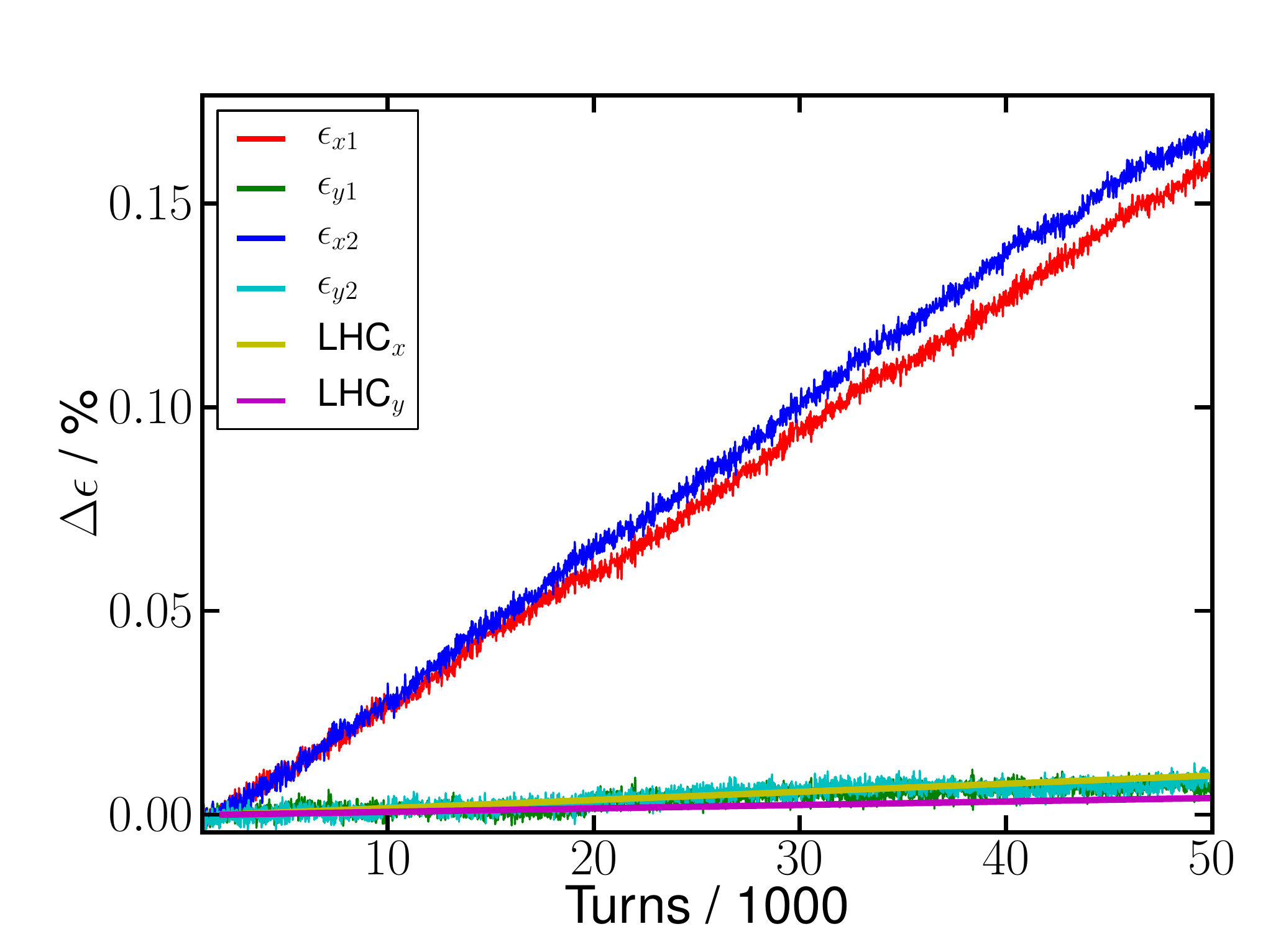}
  \caption{The emittance growth in the HL beams with damper noise (upper) and CC noise (lower).}
  \label{fig:emit_f05}
\end{figure}

Allowing for damper noise only, the emittance growth amounts to about 31 \%/h horizontally and only 4 \%/h vertically.
Compared to the numbers from 2012, the horizontal growth is considerably enhanced, while vertically a small suppression is observed.
Switching to CC noise exclusively, the horizontal emittance is blown up much more strongly, by 140 \%/h.
In the horizontal plane, the growth is still at a moderate 7 \%/h.
Figure~\ref{fig:emit_f05} visualizes the emittance as a function of time.

In order to determine how much the CCs contribute to the observed emittance growth, a simulation was run without CCs and setting the crossing angle to 0, such that the luminosity and the beam--beam parameter remained unchanged.
The resulting emittance growth agreed very well with the simulation with CCs and damper noise.
It should be pointed out, however, that chromatic effects were not included in the simulations shown here.

\begin{figure}[t]
  \centering
  \includegraphics*[width=75mm]{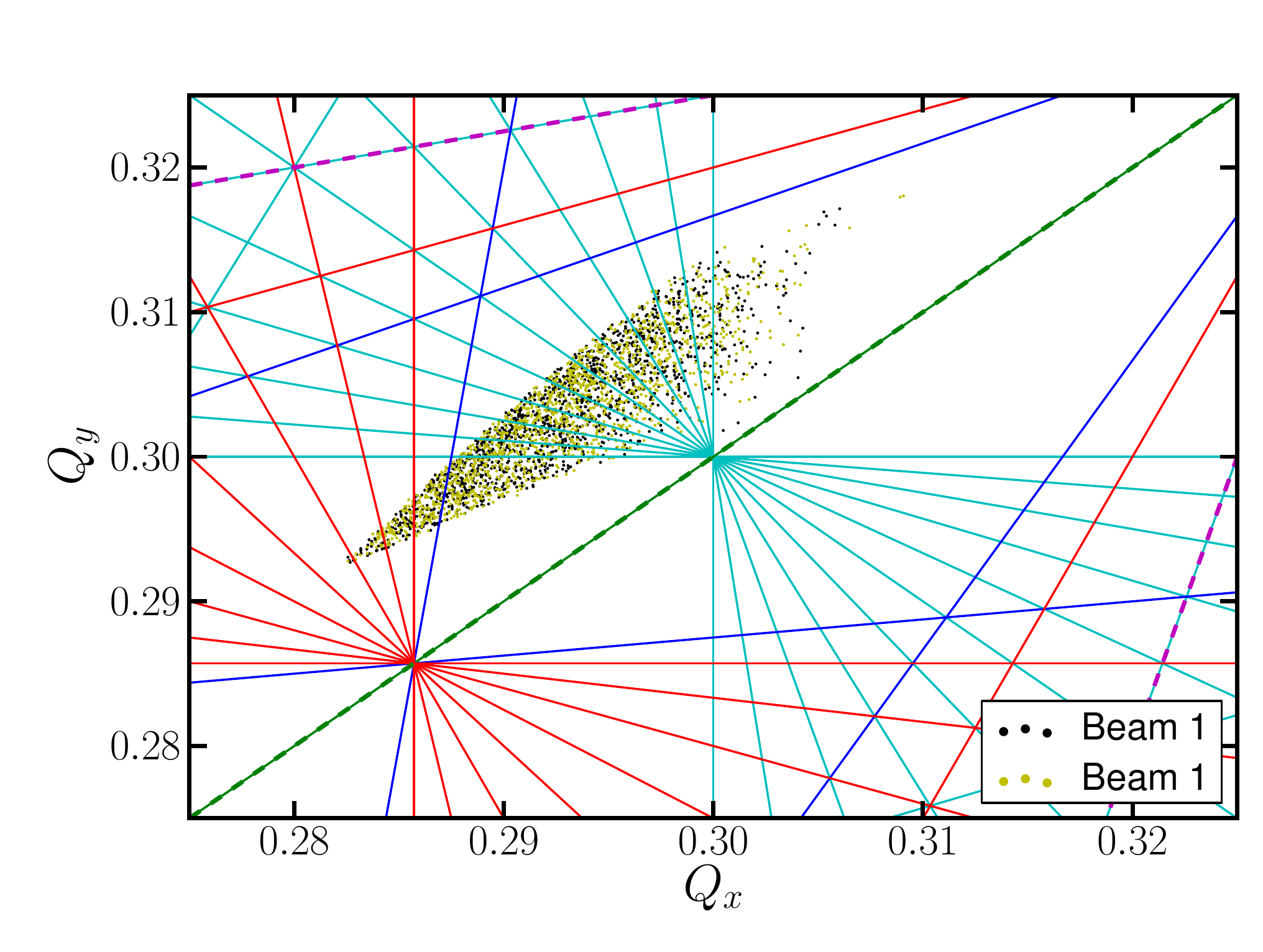}
  \caption{The tune diagram of the HL beams: red lines, seventh-order resonances; blue lines, ninth-order resonances; cyan lines, tenth-order resonances.}
  \label{fig:tune}
\end{figure}

The observed asymmetry between the two transverse planes motivated a more detailed investigation of the role of the tunes.
Exchanging the tunes of the horizontal and the vertical plane, as well as the damper parameters (which depend on the tune) led approximately to an exchange of the growth rates in the two planes.
A look at the tune diagram helps us to understand the reason.
The beam--beam force in HL collisions gives rise to a considerable tune shift and spread, spreading the tune over several seventh- and tenth-order lines and one ninth-order resonance line, as displayed in Fig.~\ref{fig:tune}.

\begin{figure}[b]
  \centering
  \includegraphics*[width=80mm]{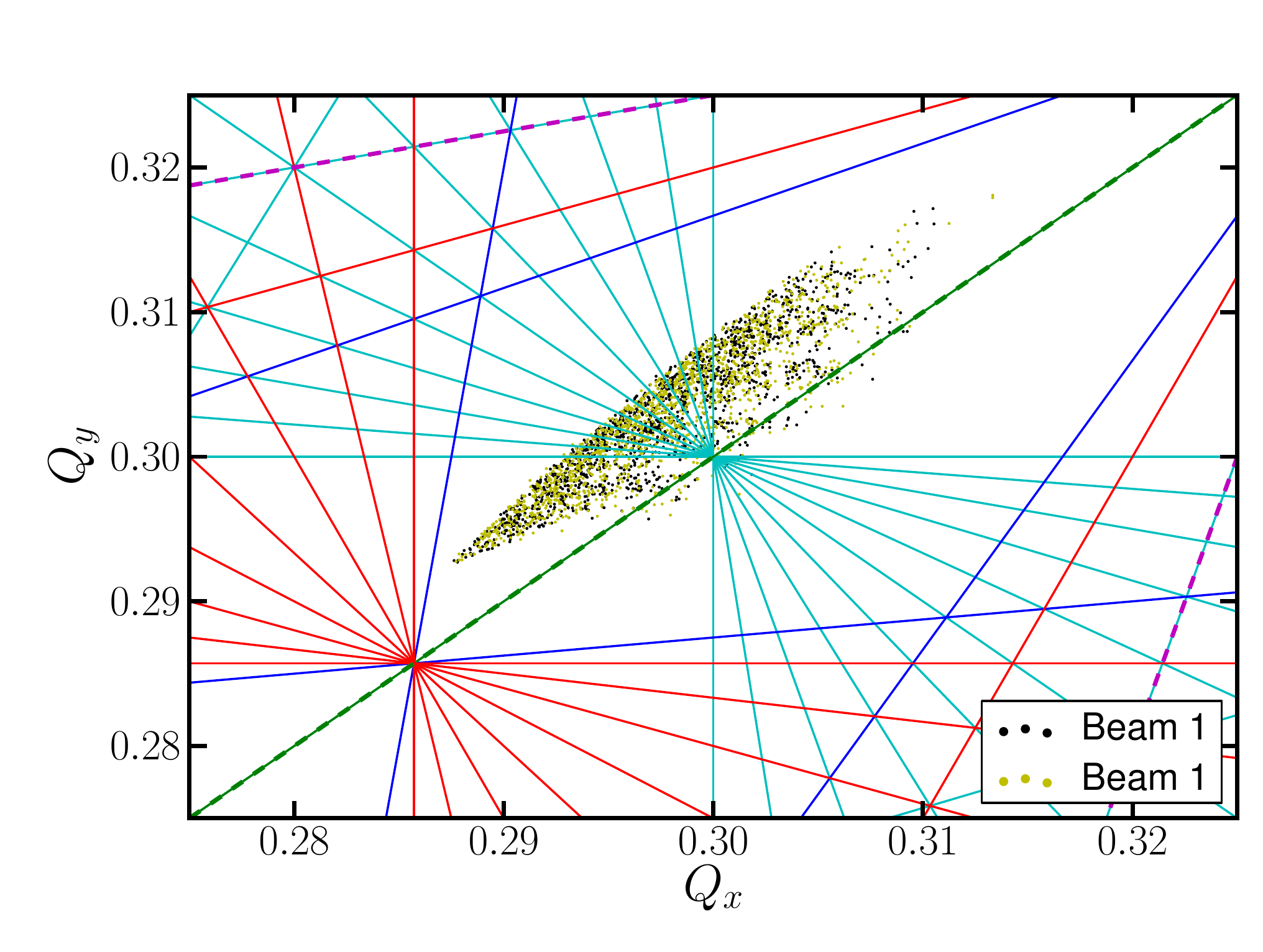}
  \caption{The tune diagram with $Q_x=64.315$. The colour code of the resonances is the same as in Fig.~\ref{fig:tune}.}
  \label{fig:tunes2}
\end{figure}

\begin{figure}
  \centering
  \includegraphics*[width=80mm]{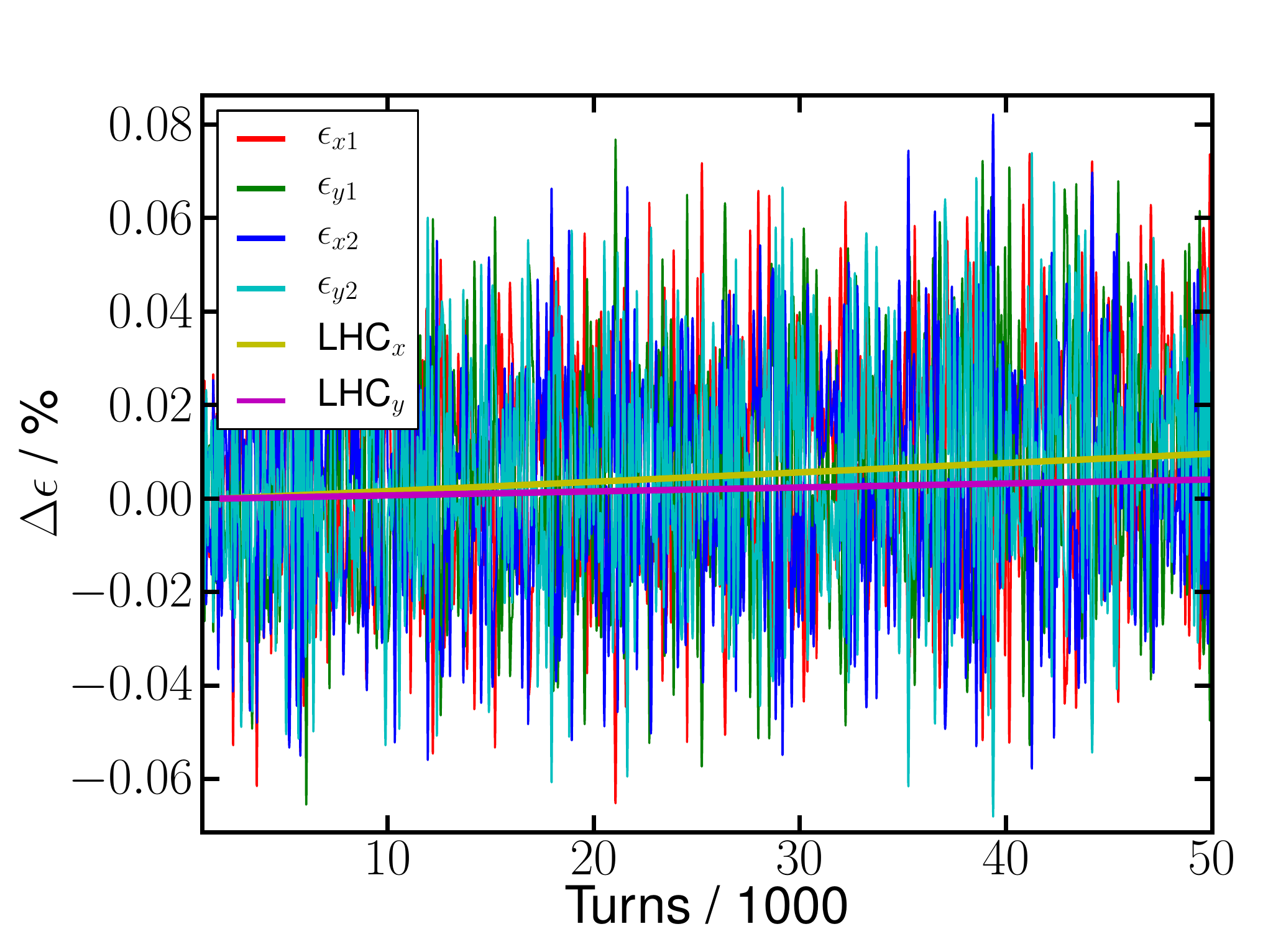}
  \includegraphics*[width=80mm]{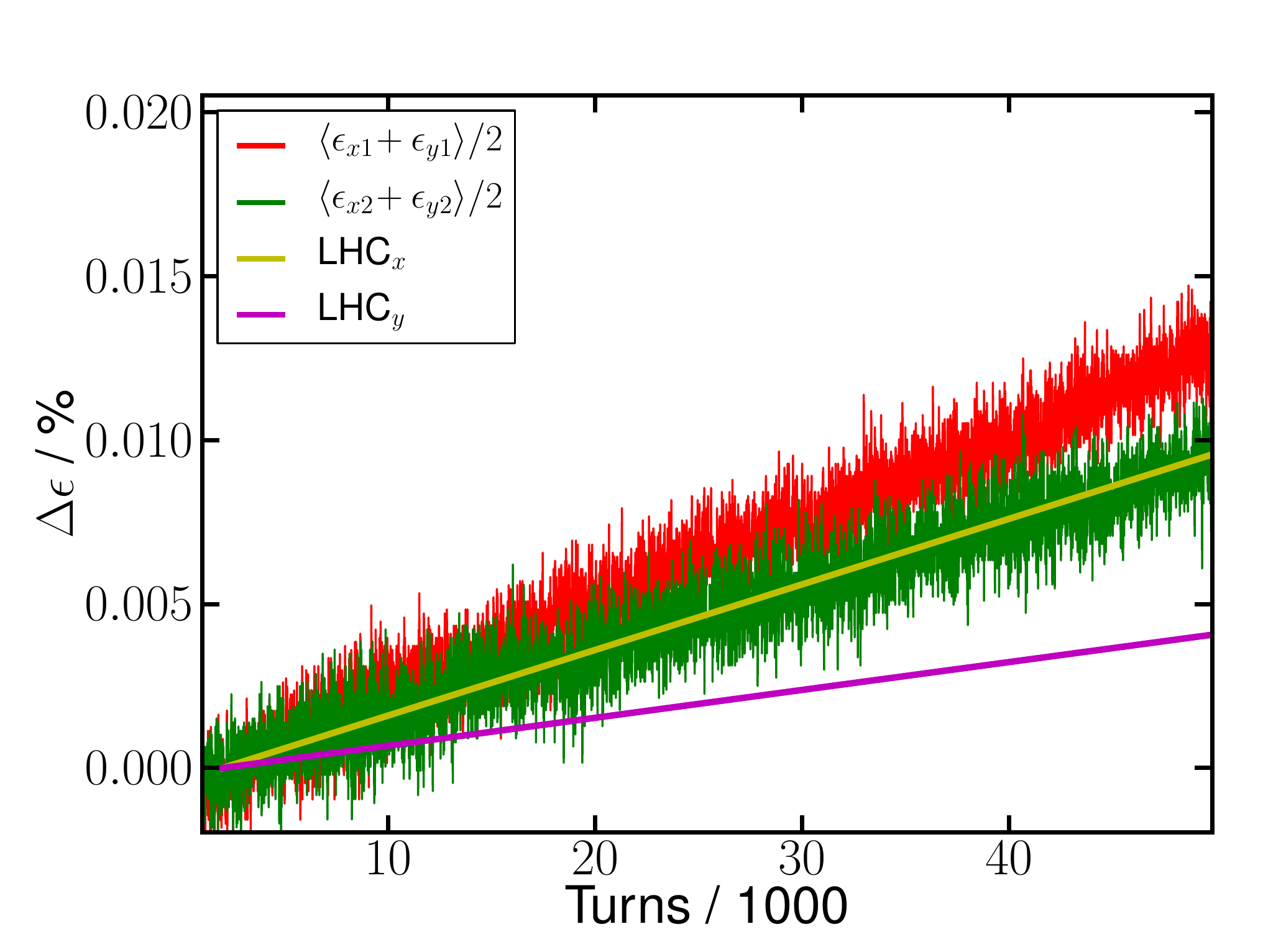}
  \caption{The emittance growth with $Q_x=64.32$ and damper noise. Upper: all four emittances. Lower: the average of the horizontal and vertical emittance for each beam. Note that the data in the upper figure have been smoothed by a Gaussian convolution.}
  \label{fig:loweps}
\end{figure}

Increasing the working point by 0.005 horizontally avoids the seventh- and ninth-order resonances, as Fig.~\ref{fig:tunes2} reveals.
A simulation with $Q_x=64.315$ was run and yielded 16 \%/h horizontal and 8 \%/h vertical emittance growth with damper noise only.
Shifting the working point further to $Q_x=64.32$ produced still better results.
The emittance growth dropped to 8 \%/h horizontally and 11 \%/h vertically after adjusting the phases of the Hilbert filter to the new working point.
The emittances for this run are shown in Fig.~\ref{fig:loweps}.
The coupling due to the equal fractional tunes leads to a rapid emittance exchange, which looks like noise.
However, as the lower part of Fig.~\ref{fig:loweps} demonstrates, the averages of the transverse emittances in both beams increase linearly to a very good approximation.

\section{Conclusion}
A simulation based on actual LHC beam parameters has allowed us to reproduce the observed emittance growth by adjusting the noise level in the transverse damper.
This noise level provides a reference for simulations of the HL-LHC.
Simulations with CCs have shown that the beam is very sensitive to white noise in the CC phase.
A significant emittance growth has been found in simulations with HL beam parameters and the nominal working point.
Increasing the horizontal tune, so as to avoid some higher-order resonances, substantially reduced the emittance growth.
Noise in the damper has only a moderate effect on the emittance.
Further studies are required to produce more accurate data.

\section{Outlook}
After gaining some insights into the emittance evolution with realistic noise in the damper, and with less realistic noise in the CC phase, correlated noise and sinusoidal perturbations are of interest.
The modelling of the beam dynamics will be refined to account for effects that have been ignored so far; for example, chromaticity.
Furthermore, the simulation parameters need to be updated as the projected parameters change over time and estimations of the performance of future hardware become more precise.
In particular, having collisions at a third IP (without CCs), which may become the regular operational scenario, will have a tremendous effect on the beam dynamics.
The strong dependence of the emittance growth on the working point motivates an optimization of the working point for the HL beams.
Also the 25\,ns bunch spacing scenario, which is preferred by the experimenters at the LHC, is going to be studied.

\section{ACKNOWLEDGEMENTS}
The authors acknowledge the support of W. H\"ofle, G. Trad, and M. Schaumann, all from CERN, by providing important information about the damper, the LHC beam parameters, and intra-beam-scattering, respectively.

\end{document}